\documentclass[twocolumn,prl,superscriptaddress,noshowpacs,epsf]{revtex4}

\usepackage{graphicx}

\begin{document}
\title{Efficient one-step generation of large cluster states 
with solid-state circuits}
\date{\today}
\author{J. Q. You}
\affiliation{Department of Physics and Surface Physics Laboratory (National Key
Laboratory), Fudan University, Shanghai 200433, China}
\affiliation{Frontier Research System, The Institute of Physical
and Chemical Research (RIKEN), Wako-shi 351-0198, Japan}
\author{Xiang-bin Wang}
\affiliation{Frontier Research System, The Institute of Physical
and Chemical Research (RIKEN), Wako-shi 351-0198, Japan}
\affiliation{CREST, Japan Science and Technology Agency (JST),
Kawaguchi, Saitama 332-0012, Japan}
\affiliation{Department of Physics, Tsinghua University, Beijing 100084, China}
\author{Tetsufumi Tanamoto}
\affiliation{Corporate R \& D Center, Toshiba Corporation, Saiwai-ku, 
Kawasaki 212-8582, Japan}
\author{Franco Nori}
\affiliation{Frontier Research System, The Institute of Physical
and Chemical Research (RIKEN), Wako-shi 351-0198, Japan}
\affiliation{Center for Theoretical Physics, Physics Department,
Center for the Study of Complex Systems,
University of
Michigan, Ann Arbor, MI 48109-1040, USA}

\begin{abstract}
Highly entangled states called cluster states are a universal resource 
for measurement-based quantum computing (QC). Here we propose an efficient method 
for producing large cluster states using superconducting quantum circuits.
We show that a large cluster state can be efficiently generated in just one step 
by turning on the inter-qubit coupling for a short time. Because 
the inter-qubit coupling is only switched on during the time interval for 
generating the cluster state, our approach is also convenient for preparing the 
initial state for each qubit and for implementing one-way QC 
via single-qubit measurements. Moreover, the cluster state is robust against 
parameter variations. 

\end{abstract}
\maketitle

Quantum computing (QC) with highly entangled states, known as cluster 
states, takes advantage of both entanglement and measurement in 
a remarkable way~\cite{RR,HJB,MAN,Lee,Duan,Dur}.
In sharp contrast to conventional QC, which uses unitary one- and two-qubit 
logic operations, this new type of QC is performed through 
only single-qubit projective measurements on a cluster state.
This measurement-based QC is termed ``one-way" because it proceeds in 
an inherently time-irreversible manner. Moreover, 
it is universal in the sense that any quantum circuit and quantum gates 
can be implemented on a suitable cluster state~\cite{RR}.

For one-way QC, the initial cluster state should be first generated. 
This highly entangled state provides a universal resource for QC. 
Ideally, it is desirable to produce 
a cluster state in just one step on a scalable circuit, so as to have 
efficient QC. However, this is challenging. 
Recently, a quantum-optics experiment~\cite{WAL} implemented one-way QC 
through local non-deterministic Bell measurements. Even though the cluster 
state was generated in one step, its generation probability was extremely low.
Moreover, it is 
hard to implement scalable QC with optical cluster states 
due to the difficulty of large-scale integration in the optical devices.   
Alternatively, solid-state QC with cluster states were proposed~\cite{Loss,Wein} 
using the Heisenberg exchange interaction between electron spins in quantum dots. 
In these approaches, additional rotations are performed on individual qubits 
in order to obtain an effective Ising-like Hamiltonian for producing the 
cluster state. Also, several steps, instead of the ideal one step, are required 
to achieve a quantum-dot cluster state. 

\begin{figure}
\includegraphics[width=3.3in,  
bbllx=23,bblly=263,bburx=570,bbury=685]{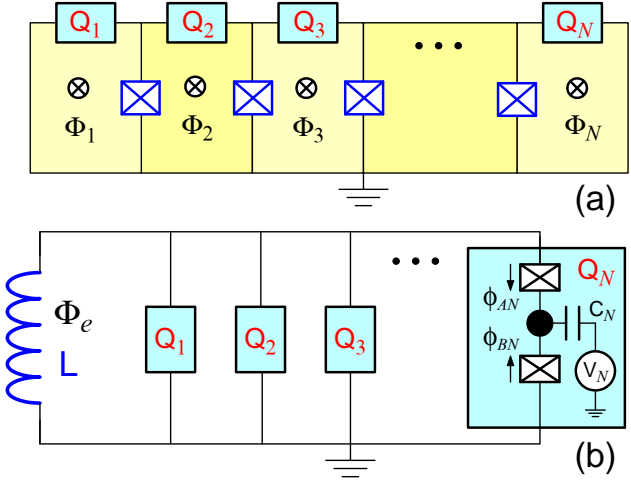} \caption{(Color online) 
Schematic diagrams of two arrays of superconducting charge qubits (blue boxes).
(a)~Qubit array ($Q_1,Q_2,\dots,Q_N$) with every nearest-neighbor charge qubits 
coupled by a large Josephson junction (JJ), shown as a crossed rectangle, 
that acts as an effective inductance $L_J$.
The inter-qubit coupling is induced by the externally applied magnetic flux 
$\Phi_i$ in each small superconducting loop (yellow) connecting qubit $Q_i$ 
and one or two large JJs. For simplicity, all large JJs are assumed to have 
the same Josephson coupling energy $E_{J0}$.
(b)~Qubit array with long-range interactions, where all charge qubits are 
connected in parallel to a common inductance $L$. The inter-qubit coupling 
is induced by the external magnetic flux $\Phi_e$ through the inductance $L$.
As a typical example, we explicitly show the schematic diagram of charge qubit 
$Q_N$, where a superconducting island (denoted as a solid circle) is connected to 
two JJs (with phase drops $\phi_{AN}$ and $\phi_{BN}$) 
and biased by a voltage $V_N$ through the gate capacitor $C_N$. }
\label{fig1}
\end{figure}

Here we propose an efficient method for one-step generation of large cluster states 
using superconducting quantum circuits. These circuits are based on 
Josephson junctions (JJs) and are regarded as promising candidates of solid-state 
qubits (see, e.g., \cite{YN}). We consider two scalable quantum circuits 
in which an {\it inductive coupling} is employed to couple nearest-neighbor 
charge qubits in one circuit [Fig.~\ref{fig1}(a)] and arbitrarily separated 
charge qubits in the other circuit [Fig.~\ref{fig1}(b)]. Both circuits give rise 
to an Ising-like Hamiltonian, but the inter-qubit interactions are 
nearest-neighbor and long-range, respectively.
 
Because the decoherence time for quantum states is limited, in order to have 
efficient one-way QC, it is essential to generate a cluster state  
in a deterministic and fast way, so that only a short time is consumed.
Also, it should be convenient to prepare the initial state for each qubit and 
to perform local single-qubit measurements. Furthermore, for a solid-state system, 
the produced cluster state should be robust against unavoidable parameter variations. 
Our proposed JJ circuits meet these requirements.
First, the cluster state is generated in just one step by turning on, for a short 
time, the inter-qubit coupling. Also, this cluster-state generation is deterministic, 
in sharp contrast to the extremely low probability of generating non-deterministic 
optical cluster states~\cite{WAL}. 
Second, because the inter-qubit coupling is turned on only when generating 
the cluster state, the preparation of the initial state for 
each qubit can be easily achieved,
before generating the cluster state, via local single-qubit operations. 
Moreover, due to the absence of inter-qubit coupling after generating the 
cluster state, local projective measurements used for one-way QC can also be 
conveniently performed via single qubits. 
Third, the cluster state is robust against unavoidable parameter variations because 
its decoherence time is not sensitive to such parameter variations.
Furthermore, all qubits work at the optimal point (namely, the degenercy point) 
where the quantum state has a longer decoherence time.

{\it Qubit array with nearest-neighbor interactions.}{\bf---}Consider 
a chain of qubits described by the Hamiltonian
\begin{equation}
H=\hbar g(t)\sum_{i,j}\Gamma(i-j)\,\frac{1\pm\sigma_x^{(i)}}{2}
\frac{1\pm\sigma_x^{(j)}}{2},
\label{E1}
\end{equation}
where $\Gamma(i-j)$ specifies the interaction range of the qubits.
Similar to the quantum Ising model used for producing cluster states~\cite{RR,HJB},
this Hamiltonian is also Ising-like, but its anisotropic direction and 
the ``magnetic" field are along the $x$ direction, instead of the usual 
$z$ direction. Below we first focus on a chain of superconducting charge qubits 
with nearest-neighbor interactions.

As shown in \cite{YTN}, two charge qubits can be coupled by a shared 
inductance. Because a JJ can behave like an effective inductance, one can also 
replace the common inductance with a large JJ~\cite{YPRB,Lantz,Wang}. 
Figure~\ref{fig1}(a) shows an array of charge qubits with a large JJ connected to 
every pair of nearest-neighbor qubits. This large JJ directly couples 
the nearest-neighbor charge qubits. Also, the non-nearest-neighbor qubits 
can be coupled via the large JJs, but the interactions are negligibly 
small. Here we use the charge states $|0\rangle$ and $|1\rangle$ 
as the basis states, which correspond to zero and one 
extra Cooper pairs in the superconducting island of each qubit. 
The Hamiltonian of the charge-qubit array can be reduced to 
$H_A=\sum_{i=1}^N [H_i+ \Lambda_{i,i+1}\sigma_x^{(i)}\sigma_x^{(i+1)}]$,
with $\Lambda_{N,N+1}=0$. The Hamiltonian $H_i$ of the $i$th charge qubit is 
$H_i=\varepsilon_i(V_i) \sigma_z^{(i)}-\overline{E}_{Ji}\sigma_x^{(i)}$,
with $\varepsilon_i(V_i)=\frac{1}{2}E_{ci}(C_iV_i/e - 1)$
and $\overline{E}_{Ji}=E_{Ji}\cos(\pi\Phi_i/\Phi_0)$. 
We assume that the charge qubit works in the charging regime 
with $E_{ci}\gg E_{Ji}$.
Here $E_{ci}$ is the charging energy of the superconducting island in 
the $i$th qubit and $E_{Ji}$ is the Josephson coupling energy of the two identical 
JJs coupled to the island; $V_i$ is the gate voltage applied to the qubit 
and $\Phi_i$ is the externally applied magnetic flux through a small loop 
connecting the $i$th qubit and one or two large JJs [see Fig.~\ref{fig1}(a)].
The flux-dependent inter-qubit coupling is given by~\cite{YPRB}
\begin{equation}
\Lambda_{i,i+1}=L_J\left(\frac{\pi^2 E_{Ji}E_{J,i+1}}{\Phi_0^2}\right)
\sin\left(\frac{\pi\Phi_i}{\Phi_0}\right)
\sin\left(\frac{\pi\Phi_{i+1}}{\Phi_0}\right),
\end{equation}
where the large JJ acts as an effective inductance $L_J=\Phi_0/2\pi I_0$, 
with $I_0=2\pi E_{J0}/\Phi_0$ being its critical current. 

When each charge qubit is shifted to work at the degeneracy point $C_iV_i/e=1$, 
the Hamiltonian of the charge-qubit array becomes 
$H_A=\sum_{i=1}^N [-\overline{E}_{Ji}\,\sigma_x^{(i)}
+\Lambda_{i,i+1}\,\sigma_x^{(i)}\sigma_x^{(i+1)}]$. 
Let $\frac{1}{2}\overline{E}_{Ji}=\Lambda_{i,i+1}\equiv \frac{1}{4}\hbar g$ 
(for $i=2,3,\dots,N-1$), and $\overline{E}_{J1}=\Lambda_{12}=\overline{E}_{JN}
\equiv \frac{1}{4}\hbar g$. These conditions can be readily  
satisfied by choosing suitable $E_{Ji}$ and $\Phi_i$, because $\overline{E}_{Ji}$ 
decreases from $E_{Ji}$ to zero and $\Lambda_{i,i+1}$ increases from zero to 
$(\pi/\Phi_0)^2 L_JE_{Ji}E_{J,i+1}$ for $0<\Phi_i/\Phi_0<\frac{1}{2}$.  
The reduced Hamitonian can be written as 
\begin{equation}
H_A=\hbar g\sum_{i=1}^{N-1}\frac{1-\sigma_x^{(i)}}{2}\frac{1-\sigma_x^{(i+1)}}{2},
\label{HA}
\end{equation} 
which has the form of Eq.~(\ref{E1}) with nearest-neighbor interactions 
$\Gamma(i-j)=\delta_{i+1,j}$.  

Initially, the external flux is not applied, so that no inter-qubit coupling 
is induced and one can manipulate each charge qubit separately.  
We first prepare all qubits in the state $|0\rangle_i$. This initial state
can be produced by applying a gate voltage to the left ($C_iV_i/e\sim 0$ ) 
of the degeneracy point and it corresponds to 
the ground state of the system. Then, shift the gate voltage $V_i$ fast to 
the degeneracy point ($C_iV_i/e=1$) and turn on the 
externally applied magnetic flux $\Phi_e$ to trigger the inter-qubit coupling 
for a period of time $t$.
The unitary transformation generated by the Hamiltonian (\ref{HA}) is 
$U(t)=\exp(-iH_A \,t/\hbar)$.
The inital state of each charge qubit can be written as 
$|0\rangle_i=(|-\rangle_i + |+\rangle_i)/\sqrt{2}$, 
where $|\pm\rangle_i=(|0\rangle_i \mp |1\rangle_i)/\sqrt{2}$ are eigenstates 
of $H_i=-\overline{E}_{Ji}\sigma_x^{(i)}$ with eigenvalues $\pm \overline{E}_{Ji}$. 
For the values $gt=(2n+1)\pi$, where $n$ is an integer, the generated 
state of the charge-qubit array is a highly entangled cluster state: 
\begin{equation}
|\phi_N\rangle=\frac{1}{2^{N/2}}\bigotimes_{i=1}^N
\,\left(|-\rangle_i + |+\rangle_i\,\sigma_x^{(i+1)}\right),
\label{CA}
\end{equation}
with the convention $\sigma_x^{(N+1)}\equiv 1$. 

The generation of cluster states in an array of {\it capacitively} coupled charge 
qubits was proposed in \cite{Tana}. 
Because of the limitation due to the capacitive inter-qubit 
interaction, the approach in \cite{Tana} is valid when each qubit works
far away from the degeneracy point. This is not desirable because 
the decoherence time of a charge qubit becomes much shorter 
away from the degeneracy point.
Furthermore, because the capacitive inter-qubit coupling is fixed~\cite{Pash}, 
it is difficult to prepare the initial state for each qubit. 
However, the generation 
of cluster states proposed here employs an array of {\it inductively} coupled 
charge qubits. This new proposal has obvious advantages: (1)~Each charge qubit 
works {\it at} the degeneracy point when generating a cluster state, 
where the qubit has a {\it longer} decoherence time; (2)~the initial state of all 
qubits can be easily prepared by turning off the external magnetic flux
and shifting the gate voltage away from the degeneracy point; 
(3)~when the initial state is prepared, the cluster state can be
readily generated by applying the external flux $\Phi_i$ for a period of time;
(4)~After generating the cluster state, no external magnetic flux is applied 
and the inter-qubit coupling is switched off. This becomes convenient 
for implementing one-way QC via local single-qubit measurements on the generated 
cluster state.

{\it Qubit array with long-range interactions.}{\bf---}When multiple charge qubits 
are connected to a commonly shared inductance [see Fig.~\ref{fig1}(b)], 
not only nearest-neighbor but also  
distant qubits can be coupled by this common inductance~\cite{YTN}. 
Because the common inductance for coupling the charge qubits has a large 
value ($L\sim 10$~nH)~\cite{YTN}, if the circuit is not too large, 
the inductances of the circuit, except $L$, can be neglected.
The reduced Hamiltonian of the system is given by
$H_B=\sum_{i=1}^N H_i -\sum_{i,j(j>i)}^N \Lambda_{ij}\sigma_x^{(i)}\sigma_x^{(j)}$. 
Here $\overline{E}_{Ji}$ in the single-qubit Hamiltonian $H_i$ becomes 
$\overline{E}_{Ji}=E_{Ji}\cos(\pi\Phi_e/\Phi_0)$, 
with $\Phi_e$ being the externally applied magnetic flux 
through the common inductance $L$.  The inter-qubit coupling is 
\begin{equation}
\Lambda_{i,j}=L\left(\frac{\pi^2 E_{Ji}E_{Jj}}{\Phi_0^2}\right)
\sin^2\left(\frac{\pi\Phi_e}{\Phi_0}\right).
\end{equation}
Let $\overline{E}_{Ji}/(N-1)=\Lambda_{ij}\equiv \frac{1}{4}\hbar g$, 
for $1\leq i,j \leq N$ and $j>i$. 
This condition can be satisfied using $N$ identical charge qubits and
a suitable $\Phi_e$. While fulfilling this condition and simultaneously having 
each charge qubit work {\it at} the degeneracy point, the Hamiltonian becomes  
\begin{equation}
H_B=-\hbar g\sum_{i,j (j>i)}^N \frac{1+\sigma_x^{(i)}}{2}\frac{1+\sigma_x^{(j)}}{2},
\label{HB}
\end{equation}
which corresponds to Eq.~(\ref{E1}) with long-range interactions.

The initial state of each charge qubit, 
$|0\rangle_i=(|-\rangle_i + |+\rangle_i)/\sqrt{2}$,
is also prepared by both turning off the external flux $\Phi_e$ and  
applying a gate voltage to the left of the degeneracy point.
Furthermore, we shift the gate voltage fast to the degeneracy point and 
apply the flux $\Phi_e$ for a period of time $t$. 
The unitary transformation given by the Hamiltonian (\ref{HB}) is 
$U(t)=\exp(-iH_B \,t/\hbar)$.
At $gt=(2n+1)\pi$, the generated cluster state is  
\begin{equation}
|\psi_N\rangle=\frac{1}{2^{N/2}}\bigotimes_{i=1}^N
\,\left(|-\rangle_i \,(-1)^{N-i}\!\!\!\prod_{j=i+1}^N \!\sigma_x^{(j)} 
+ |+\rangle_i\right), 
\label{CB}
\end{equation}
which is also a highly entangled state. In Eq.~(\ref{CA}), 
the operator $\sigma_x^{(i+1)}$ acts on the states $|\pm\rangle$ 
of the $(i\!+\!1)$th qubit. However, for the cluster state $|\psi_N\rangle$, 
the operator $\sigma_x^{(j)}$ acts on the states $|\pm\rangle$ of the qubits 
$j=i+1,\dots,N$, with $i=1,2,\dots,N-1$; this is due to the long-range nature 
of the inter-qubit coupling in Hamiltonian (\ref{HB}).   

{\it Parameter variations and robustness of cluster states.}{\bf---}As in other 
solid-state systems, parameter variations unavoidably occur when fabricating 
JJ circuits. When the parameters $E_{Ji}$ vary in the charge-qubit array 
in Fig.~\ref{fig1}(a), because $\overline{E}_{Ji}=E_{Ji}\cos(\pi\Phi_i/\Phi_0)$, 
the condition $\overline{E}_{J1}=\overline{E}_{JN}=
\frac{1}{2}\overline{E}_{Ji}=\frac{1}{4}\hbar g$, with $i=2,3,\dots,N-1$,
can be satisfied by adjusting the local magnetic 
fluxes $\Phi_i$. However, if $L_J$ and $E_{Ji}$ vary, the condition 
$\Lambda_{i,i+1}=\frac{1}{4}\hbar g$ cannot be satisfied. 
In order to fulfill this condtion, 
one can connect a current source in parallel to each large JJ and bias 
the JJ with a current $I_{bi}<I_0$. Now the inter-qubit coupling becomes 
\begin{eqnarray}
\Lambda_{i,i+1}\!&\!=\!&\!L_{Ji}\left(\frac{\pi^2 E_{Ji}E_{J,i+1}}{\Phi_0^2}\right)
\sin\left(\frac{\pi\Phi_i}{\Phi_0}+\frac{1}{2}\gamma_i\right) \nonumber \\
&&\!\times 
\sin\left(\frac{\pi\Phi_{i+1}}{\Phi_0}-\frac{1}{2}\gamma_i\right),
\end{eqnarray}
where the effective inductance for each large JJ is $L_{Ji}=\Phi_0/2
\pi I_0\cos\gamma_{i}$, with $\gamma_{i}=\sin^{-1}(I_{bi}/I_0)$.
Here the condition $\Lambda_{i,i+1}=\frac{1}{4}\hbar g$ can be readily 
satisfied by changing the bias current $I_{bi}$.  
As for the charge-qubit array in Fig.~\ref{fig1}(b) and the capacitively 
coupled charge qubits~\cite{Tana}, the conditions for obtaining 
an Ising-like Hamiltonian cannot be fully satisfied for varying qubit parameters. 
Therefore, the charge-qubit circuit in Fig.~\ref{fig1}(a) should be 
advantageous for suppressing the effects of unavoidable parameter variations. 

Below we further show the robustness of the cluster states against parameter 
variations. According to the Fermi golden rule, the relaxation rate of
the $i$th charge qubit is $\Gamma^{(i)}_1\equiv 1/T^{(i)}_1
=\frac{1}{2}A_i S_i(\Omega)$, 
where $A_i=\overline{E}_{Ji}^2/(\varepsilon_i^2+\overline{E}_{Ji}^2)$ 
and $S_i(\omega)$ is the power spectrum of the charge noise dominant 
in the charge qubit. For a typical Gaussian 
noise~\cite{Naka}, the dephasing factor is   
$\eta_i(\tau)=B_i\int d\omega S_i(\omega)\frac{\sin^2(\omega\tau/2)}
{2\pi(\omega/2)^2}$, 
with $B_i=\varepsilon_i^2/(\varepsilon_i^2+\overline{E}_{Ji}^2)$. 
The dephasing rate $\Gamma^{(i)}_{\varphi}\equiv 1/T^{(i)}_{\varphi}$ is defined 
by $\eta_i(T^{(i)}_{\varphi})=1$.  
Following the Bloch-Redfield theory (see, e.g., \cite{Cohen}), 
the decoherence rate $\Gamma^{(i)}_2\equiv 1/T^{(i)}_2$ is  
$\Gamma^{(i)}_2=\frac{1}{2}\Gamma^{(i)}_1+\Gamma^{(i)}_{\varphi}$. 
Because all inter-qubit couplings are switched off after generating a cluster state,
the decoherence time $T_2$ of the cluster state is given by
$1/T_2=\sum_i 1/T^{(i)}_2$. 
Here all charge qubits work at the degeneracy point 
$\varepsilon_i\approx 0$, thus $A_i\approx 1-(\varepsilon_i/\overline{E}_{Ji})^2$ 
and $B_i\approx (\varepsilon_i/\overline{E}_{Ji})^2$. Obviously, 
$A_i$ and $B_i$ are weakly affected by the variations of the parameters $E_{Ji}$. 
This indicates that the decoherence time $T_2$ of the cluster state is not 
sensitive to the parameter variations. Therefore, the cluster state 
is robust against the unavoidable parameter variations.

{\it Discussion and conclusion.}{\bf---}Each Josephson coupling energy 
for a charge qubit is typically $E_J/h \sim 10$~GHz 
(see, e.g., \cite{Pash}), which corresponds to a switching time 
$\tau_1\sim 0.1$~ns for the single-qubit operation. 
For the charge-qubit array in Fig.~\ref{fig1}(a), the inter-qubit coupling is
$\Lambda\sim L_J(\pi E_J/\Phi_0)^2$, 
where $L_J=\Phi_0/2\pi I_0=(1/E_{J0})(\Phi_0/2\pi)^2$. 
Choosing, e.g., $E_{J0}=5E_J$, one obtains
$\Lambda/h\sim 0.5$~GHz. Because $\hbar g/4=\Lambda$, the shortest time 
to generate the cluster state is $t_s=\pi/g\sim 0.25$~ns, comparable to 
the switching time $\tau_1$ of the single-qubit operation. For the array of 
charge qubits coupled by a common inductance $L$ [see Fig.~\ref{fig1}(b)],  
the inter-qubit coupling is $\Lambda\sim L(\pi E_J/\Phi_0)^2$. 
Using $L=10$~nH, one has $\Lambda/h\sim 1.1$~GHz. The corresponding 
shortest time for generating the cluster state is 
$t_s\sim 0.11$~ns $\approx\tau_1$.
Let $T_2$ be the decoherence time of a qubit. The decoherence time of 
$N$ weakly coupled qubits can be estimated as $T_2^{(N)}\sim T_2/N$. 
For a charge qubit with $T_2\sim 0.5$~$\mu$s at the degeneracy point~\cite{Yale}, 
considering an array with $N=100$ charge qubits, one obtains
$T_2^{(N)}\sim 5$~ns. This decoherence time is longer than the shortest time $t_s$ 
for generating the cluster state.
Here the common inductance is chosen to be large (e.g., $L=10$~nH), but the inter-qubit
couplings are still weak, and those couplings are turned off after 
generating a cluster state. Thus, the effects of $L$ on the decoherence of the
cluster state are small. Moreover, these decoherence effects can be further reduced 
when $L$ is replaced by a large JJ acting as an effective inductance. 

For the usual quantum Ising model, where its anisotropic direction and 
the ``magnetic" field are both along the $z$ direction, the basis states used for 
representing a cluster state are the eigenstates $|0\rangle_i$ and $|1\rangle_i$ 
of $\sigma_z^{(i)}$. To implement one-way QC, single-qubit projective measurements 
are performed~\cite{RR} on the basis states 
$|\pm\rangle_i$, namely, the eigenstates of $\sigma_x^{(i)}$. 
In our proposed charge-qubit arrays 
the reduced 
Hamiltonian is also Ising-like, but its anisotropic direction and the ``magnetic" 
field are along the $x$ direction, instead of the $z$ direction~\cite{RR}. 
Now the cluster state is represented using 
basis states $|\pm\rangle_i$, instead of $|0\rangle_i$ and $|1\rangle_i$. 
Correspondingly, the single-qubit projective measurements are performed 
on the basis states $|0\rangle_i$ and $|1\rangle_i$. In charge qubits, 
these states correspond to zero and one extra Cooper pairs in the 
superconducting island of each qubit. The local single-qubit projective 
measurements on these basis states can be implemented using, e.g., either a probe 
junction~\cite{Pash} connected to the superconducting island or 
a single-electron transistor~\cite{Asta} coupled to the charge qubit.  
Because the single-electron transistor has high efficiency for reading out the 
quantum state, it is more advantageous to use it for performing local single-qubit 
projective measurements. 
Also, because such a transistor is coupled to each charge qubit via a small 
capacitance, it only produces weak backaction on the qubit state in the absence 
of quantum measurement.     

In conclusion, we propose an efficient method for producing large cluster 
states using superconducting circuits. 
We consider two charge-qubit arrays where either nearest-neighbor or
arbitrarily separated qubits are inductively coupled. 
The initial cluster state can be efficiently generated in just one step 
by turning on the inter-qubit coupling for a short time. Also, 
our approach is convenient for preparing the initial state of the system 
and for implementing one-way QC via single-qubit measurements on the cluster state
because the inter-qubit coupling is switched off in both cases.

We would like to thank L.F. Wei for discussions. 
This work was supported in part by 
the NSA, LPS, ARO and NSF. 
J.Q.Y. was supported by the NSFC
grant Nos.~10474013 and 10534060.

\end{document}